\documentclass[useAMS,usenatbib]{mn2e}
\usepackage{graphicx}
\usepackage{amsmath}
\usepackage{amssymb}
\usepackage{xcolor}
\usepackage{hyperref}
\providecommand{\adsurl}[1]{\href{#1}{ADS}}
 
\usepackage{url}

\title{Nonextensive Statistical Analysis of Meteor Showers and Lunar Flashes}
\author[A. S. Betzler and E. P. Borges]%
{A. S. Betzler$^{1,2}$\thanks{E-mail: betzler@ufrb.edu.br (ASB)} 
and 
E. P. Borges$^{3}$\thanks{ernesto@ufba.br (EPB)}\\
$^{1}$Centro de Forma\c{c}\~ao
de Professores, Universidade Federal do Rec\^oncavo da Bahia, Amargosa, 45300-000, Brazil\\
$^{2}$Escola Polit\'{e}cnica, Universidade Federal da Bahia, Salvador,
40210-630, Brazil\\
$^{3}$Instituto de F\'{\i}sica, Universidade Federal da Bahia,
Salvador, 40170-115, Brazil\\}

\date{\empty}

\begin{document}
\maketitle

\begin{abstract}

The distribution of meteor magnitudes is usually supposed to be 
described by power-laws. However, this relationship 
is not able to model the whole data set, and the parameters are considered 
to be dependent on the magnitude intervals. 
We adopt a statistical distribution derived from Tsallis nonextensive 
statistical mechanics which is able to model the whole magnitude range. 
We combined meteor data from various sources, 
ranging from telescopic meteors to lunar impactors.
Our analysis shows that the probability distribution of magnitudes of IMO 
and MORP data are similar.
The distribution of IMO visual magnitudes indicates that $2.4\pm0.5\%$ 
of the meteors of a shower may be telescopic ($m> 6$). 
We note that the distribution of duration of lunar flashes
follows a power-law, and a comparison with the distribution of meteor showers 
suggests the occurrence of observational bias. 
The IMO sporadic meteor distribution also seems to be influenced 
by observational factors.

\end{abstract}

\begin{keywords}
meteors, methods: statistical  
\end{keywords}

\section{Introduction}
Meteor showers are phenomena resulting from the collision of meteoroid
streams with the Earth's atmosphere.
These particles originated from comets and asteroids, 
generically known as parent-bodies.
The first dynamic association between a comet and a meteor shower was
established in 1867 involving the Leonids and 55P/Tempel-Tuttle 
\citep{manson1995}. 
A similar connection between the asteroid (3200) Phaethon and the Geminids 
meteor shower was established more than a century later \citep{whipple1983}.
Studies of the physical properties of meteor showers began in the second half 
of the nineteenth century. These studies were focused on determining 
the chemical properties and brightness of meteors \citep{milman1980}.
The mass cumulative distribution of a meteor shower is usually assumed 
to be represented by an exponential function \citep{zolensky2006},
\begin{equation}
\label{eq:cummulativemass}
 \log N=A \log M + B,
\end{equation}
where 
$M$ is the meteoroid mass, 
$A$ is the integrated mass index, 
$B$ is a constant, 
and
$N$ is the meteor flux (number of events per area per time)
with mass equal or greater than $M$. 
The cumulative distribution of meteor magnitudes $m$ is given by 
\cite{baggaley1977}
\begin{equation}
\label{cumdistmagmet}
 N_m=Cr^m,
\end{equation}
where $N_m$ is the number of meteors brighter than, or equal to, $m$, 
$r$ is the ratio between the number of meteors with magnitudes
$m$ and $m+1$, 
and $C$ is a constant. 
Equations (\ref{eq:cummulativemass})  and (\ref{cumdistmagmet})  
do not properly model all the mass distribution and magnitudes 
of a meteor shower. 
This is particularly significant for meteors with $m\leq -3$.
Specifically, there is a variation of the mass index for these magnitudes. 
The use of Equation\ (\ref{eq:cummulativemass}) can infer an incorrect 
estimate of the mass flow incident on the Earth. 
An empirical solution usually adopted for this problem is to use 
Equation\ (\ref{eq:cummulativemass}) for specific mass intervals,
instead of assuming that it is valid for the whole distribution.
A fragmentation process can occur during formation, 
in the transit of meteoroids between the parent-body and the Earth,
or its interaction with the Earth's atmosphere 
(\citealt{toth2005}; \citealt{ceplecha2005};  \citealt{jenniskens2007}).
The frequency of meteoric bombardment on the Moon
is presumably similar to that on the Earth.  
This seems quite reasonable, given the size of a few million kilometres  
of meteoroid streams associated with annual showers \citep{nakamura2000}. 
The visualisation of the collision of meteoroids with the Moon is an unusual 
phenomenon. The only known records, until the 1990s, 
were associated with more energetic events than those 
commonly observed on Earth.
An example is the photographically recorded flash by \cite{stuart1956}.
Based on the obtained image, \cite{buratti2003} estimated that a
body of 20~m of diameter crashed on the Moon, 
creating a 1.5~km diameter crater.
The detection of lunar impacts was not possible until the introduction
of CCD's cameras into this type of observation, due to the short duration
and low brightness of the events \citep{ortiz2002}.
The first detection of lunar flashes was obtained by \cite{bellot2000}
and \cite{dunham2000} for the Leonids meteor shower.
The impacts can be associated with bodies with a mass on order of 
a few kilograms.
The detection of these events is subject to a combination of factors such as 
moon phase, instruments used, and the nature of the meteoroid.
This ensemble of physical and instrumental factors is called 
observational bias. 
Observational bias is particularly important in the study of
meteors and related phenomena \citep{hawkes1986}.
As for the parameters associated with lunar flashes, \cite{tost2006}
concluded that most impacts have a typical duration 
and magnitude from 0.1 to 0.5~s and $m \sim 10$ respectively. 
\cite{sanford2009} developed a model for the plume of steam due to 
an impact on the Moon, which 
suggests that parameters such as 
brightness and duration of the event are proportional to the mass 
and kinetic energy of the meteoroid. 
\cite{bouley2012} found a direct correlation between the magnitude 
and duration of the flashes. 
However, this relationship does not appear to be valid for the Leonids.
The difference between Leonids and other meteor showers is 
likely explained by the higher velocity of this swarm.
Models that describe distributions associated with
lunar flashes are not yet available in the literature.
Such models are important for the identification
of the observational bias, and to verify whether the
incidence of meteoroid streams on the Moon 
are consistent with those observed on the Earth.
We use nonextensive distributions, that emerge from Tsallis statistics
(\citealt{tsallis1988}, \citeyear{tsallis2009}) to analyse data from 
meteor surveys and lunar flashes.
In particular, we verify
that the nonextensive distribution proposed by \cite{sotolongo2007}
models quite well the cumulative magnitude distribution  
of meteor showers.
For the lunar impacts, the duration of the steam plumes is modelled 
by a power-law distribution.
The correlation between the predictions in both data sets suggests 
that there is a significant difference in the distribution of sporadic meteors
and those associated with a shower. 
Similarity in the distributions suggests a uniformity of the physical processes 
that govern the fragmentation of meteoroids and the formation of lunar flashes. 
The presence of observational bias is suggested in
the data of meteors showers and lunar flashes.

\section{Nonextensive Distributions of Meteor Showers and Lunar Flashes}

Boltzmann-Gibbs (BG) statistical mechanics needs the assumption 
(among others) of short-range interaction between particles, 
that implies strong chaos at the microscopic level
and eventually results in the ergodic hypothesis.
Distributions that emerge from BG formalism are based on the 
exponential function, that has rapidly decaying tails,
and thus events far from the expectation values are utterly rare.
Long-range gravitational interaction does not satisfy this basic assumption,
and application of the statistical mechanics formalism 
to self-gravitating systems results in difficulties and/or inadequacies,
\textit{e.g.}, negative specific heat, non-equivalence of ensembles
(\citealt{1968MNRAS.138..495L,1970ZPhy..235..339T,landsberg-1990,padna2}).
These inadequacies have been at least partially overcome 
by the use of nonextensive statistical mechanics,
that is a generalisation of BG formalism
(i.e., BG statistics is a particular case of nonextensive statistics).
Nonextensive distributions present tails that are asymptotically power-laws,
which implies that rare events, though unlikely, 
do not have vanishingly small probabilities.
The generalisation is governed by the entropic index $q$,
that controls the power-law regime of the tails;
the particular value $q=1$ recovers the usual BG formalism.
Non-Boltzmannian distributions, negative specific heat,
and compatibility with nonextensive statistical mechanics
have been computationally verified
(\citealt{latora-rapisarda-tsallis-2001,epb-tsallis-2002,tsallis-rapisarda-pluchino-borges-2007}).
Examples linking nonextensivity to astrophysical systems 
are galore. 
We point out three instances that ranges from small bodies 
in the Solar System, spherical galaxies, and the whole universe:
nonextensive distributions in asteroid rotation periods and diameters
have been observed (\citealt{betzler-borges-2012}).
Models for distribution function for spherical galaxies have been proposed, 
and density profiles have been shown to be a nonextensive equilibrium 
configuration, that can be used to describe dark matter haloes. 
The M33 rotation curve has been successfully fitted within this context
(\citealt{2011MNRAS.414.2265C}). 
Small deviations from Gaussianity of the cosmic microwave background 
temperature fluctuations have also been verified, 
and nonextensive distributions seem to properly describe the data 
by the Wilkinson microwave anisotropy probe (\citealt{2006PhLA..356..426B}).
More examples of applications of nonextensive statistical mechanics
in astrophysical systems may be found in \cite{tsallis2009}.

Fragmentation models have benefited from developments in
material science, combustion technology and geology, among others.
Some attempts to obtain the size distribution function were made using
the principle of maximum entropy (\citealt{lin1987}, \citealp{sotolongo1998}).
The expression for the BG entropy, 
in its continuous representation, is given by
\begin{equation}
\label{entropiaBG}
 S = - k\int p(x)\ln p(x)\,dx,
\end{equation}
where $p(x)$  is the probability of finding the system in a specific
microscopic state, generically represented by the dimensionless coordinate $x$, 
and $k$ is the Boltzmann's constant, that guarantees dimensional consistency 
to the expression -  we may take here $ k \equiv 1$ without loss of generality.
BG entropy is valid when short-range interactions are present.
A fragmented object may be considered as a collection of parts that
have entropy larger than the original \citep{sotolongo2007}
{($S (\cup A_i) > \sum_i  S (A_i)$).}
This suggests that it may be necessary to 
use a nonadditive entropy, rather than the additive BG entropy.
Tsallis nonadditive entropy for the distribution of mass fragments 
is written as
\begin{equation}
\label{entropiatsallis}
 S_q=k\frac{ 1- \int_{0}^{\infty} {p^q}(M)\,dM}{q-1},
\end{equation}
where $M$ is a dimensionless mass,
$q$ is the entropic index, and $p(M)$ is the probability density
of a fragment to have mass between $M$ and $M + dM$.
Maximisation of Equation (\ref{entropiatsallis}), 
constrained to the normalisation condition
\begin{equation}
\label{condicao_1}
 \int_{0}^{\infty} p(M)\,dM=1,
\end{equation}
and to the $q$-expectation value of the mass
\begin{equation}
\label{condicao_2}
 \int_{0}^{\infty} M{p^q}(M)\,dM= \langle M \rangle _q,
\end{equation}
$\langle M \rangle _q$ is the generalised $q$-expectation value, according to
\cite{curado1991} (see also \cite{tsallis1998} for
a broader view of the generalisation of the constraints) 
leads to the mass probability density of a meteor shower
\begin{equation}
\label{densprobmass}
 p(M)dM = a(1+bM)^{-\frac{1}{q-1}}dM.
\end{equation}
Normalisation condition, Equation\ (\ref{condicao_1}), 
imposes that $ a=b(2 -q)/(q-1)$.
The inverse cumulative distribution
$P_\geq (M) = N_\geq (M)/N_t =\int_M^\infty p(M')\,\mathrm{d}M',
N_\geq(M)$
is the cumulative number of meteors with mass equal to
or greater than $M$, and $ N_t$ is the total number of meteors,
leads to
\begin{equation}
\label{eq:densacumul}
 N_\geq(M) = N_t(1+bM)^{\frac{2-q}{1-q}} = N_t[\exp_q(-\beta M)]^{2-q}
\end{equation}
$(b = (q-1)\beta)$, that is identified with a $q$-exponential, defined as
\begin{equation}
\label{eq:q_exp}
 \exp_q x \equiv [1+(1-q) x]^{\frac{1}{1-q}},
\end{equation}
if $[1+(1-q) x]>0$, and $\exp_q x \equiv 0$ if $[1+(1-q) x]\le0$
\citep{tsallis-quimicanova1994}. 
This expression provides, in the asymptotic limit, a power-law regime
\begin{equation} 
\label{powerlaw}
 N(M)\sim M^{-n} 
\end{equation} 
with $n= \frac{2-q}{q-1}$, similar to Equation\ (\ref{cumdistmagmet}).
Once the relation between the magnitude and meteor mass can be expressed by an
exponential function (\citealt{jacchia1965}),
we adopt the expression used by \cite{sotolongo2007}
\begin{equation}
\label{massmagrela}
 M= M_0{e^{-\gamma m}},
\end{equation}
where $M_0=M(m=0)$, and $\gamma$ is a constant.
Following their lines, taking into account that
\begin{equation}
\label{equivalenciamagmass}
 p(m)dm=p(M)dM, 
\end{equation}
and finally integrating between $m$ and $-\infty$, to obtain the inverse 
cumulative distribution, we get
\begin{equation}
\label{qexpmag}
  N_\geq(m) = N_t [\exp_{q}(-\beta_m e^{-\gamma m})]^{2-q}
\end{equation}
($\beta_m = \beta M_0$; $\beta_m$ is taken as a fitting parameter in our 
procedure). 
The power of a $q$-exponential can easily be rewritten as another 
$q$-exponential, with a different index $q'$:
\begin{equation}
\label{qexpum}
  N_\geq(m) = N_t \exp_{q'}(-\beta'_m e^{-\gamma m})
\end{equation}
with $\frac{1}{1-{q'}}$=$\frac{2-q}{q-1}$ and $\beta'_m=(2-q)\beta_m$.
Graphical representation of Eq.\ (\ref{qexpum}) with the ordinate in logarithm 
scale ($\log$ with base 10), and the abscissa (magnitude) in linear scale, as 
shown in the main panel of Figures \ref{fig:per_2000}--\ref{fig:cap-spo},
presents two main regimes. For highly negative magnitudes (i.e., bright 
objects), Eq.\ (\ref{qexpum}) displays an ascending straight line with slope 
given by $\gamma/(q'-1) \log_{10} e$ (the rare events region).
For high values of magnitudes, the cumulative distribution tends to its upper 
value $N_t$, and the semi-log plot exhibits a quasi-flat region. Elongation of 
the ascending straight line of the rare events region
intercepts the horizontal line of the quasi-flat region at the transition point 
between regimes, $m^*=\frac{1}{\gamma} \ln[(q'-1)\beta'_m]$.
A similar twofold behaviour occurs regarding the decreasing $q$-exponential 
distribution, as explained in \cite{betzler-borges-2012}, 
and illustrated by its Fig.\ 1.
We have gathered data from meteor showers, and fitted Equation\ (\ref{qexpum})
(it is to be shown in the following sections). 

\section{Observational Data}
The cumulative distributions of meteor showers were obtained from 
the counting
of meteors in the range of visual magnitude, 
provided by the International Meteor Organization (IMO) (available at the VMDB 
-- Visual Meteor Database, 
{\tt http://www.imo.net/data/visual).
}
In this analysis, we consider meteor magnitudes ranging from $-6$ to $+6$ 
with an interval of one magnitude between the classes.  
The analysed counting was originated from observers that rated 
the local sky with limiting magnitude $ lm \ge 5.5$. 
This selection criterion has been adopted in similar studies 
(\citealt{brown1996}, \citealt{arlt2006}), 
and it aims to minimise observational bias.  
We analysed 10 showers: Geminids (GEM), Orionids (ORI), Quadrantids (QUA), 
Eta Aquariids (ETA), Lyrids (LYR), Capricornids (CAP), 
Leonids (LEO), Perseids (PER), 
Alpha Monocerotids (AMO), Southern Taurids (STA) 
and also sporadic meteors (SPO). 
The choice of these showers is associated with the variety of physical
and dynamic characteristics of each meteoroid stream,
which is possibly associated with the differences in 
their parent bodies. 
In order to verify whether the parameters of Equation\ (\ref{qexpum}) 
have a temporal variation,
we have analysed the whole observed meteors in the years
2000, 2002, 2004, 2006, 2008 and 2010 (data from VMDB) 
according to the limit magnitude criterion above mentioned.
Additionally, for the LEO, we have analysed the distribution of meteors 
in the 1999's outburst. 
To verify if any trends detected within 
the VMDB data are valid, we analysed
panchromatic magnitudes of fireballs, recorded by the Meteor Observation and
Recovery Project (MORP) survey \citep{halliday1996}.
From this data set, we studied the SPO, PER and STA fireballs. 
Specifically, the PER and STA have presented
the largest number of observations in the showers' database. 
To demonstrate the possible occurrence of systematic errors between visual,
photographic and TV data, we compared
the VMDB and MORP distributions with those detected by a TV camera installed 
in Salvador (SSA, Bahia, Brazil). 
The camera uses a CCD 1/3 inch Sony Super HAD EX View. 
This instrument is pointed towards the zenith,
and has a field of 89 degrees. 
The 49 SPO meteors detected were observed between
July and October, 2011. The meteor magnitudes were estimated using the
method of \cite{koten1999}.
With the same purpose, the SPO MORP magnitude distribution have been
compared with the Fireball Database Center (FIDAC) data set \citep{knofel1988} 
obtained in the years 1993, 1994, 1995, 1996 and 1997,
taken from ``Fireball Reports 1993--1996''
(available at {\tt http://www.imo.net/fireball/reports}).

We only considered the highest luminosity of detected meteors 
from the MORP and SSA data.
This procedure was adopted to allow a comparison with VMDB and 
FIDAC visual data. 
Visual observers generally record only the brightness peak of a meteor 
\citep{beech2007}.
The duration  of lunar flashes were collected 
by the Automated Lunar and Meteor Observatory (ALaMO)
of NASA's Meteoroid Environment Office. 
The analysed data were obtained between 2005 and 2010. 
Instrumental characteristics of this initiative
are presented in \cite{suggs2008}.
We obtained the distribution of the duration of lunar flashes 
of SPO and LEO, GEM, LYR, Taurids (TAU), ORI and QUA showers. 
The duration of the flashes were obtained by multiplying the number of frames 
in which the events are recorded by $1/30$~s, that corresponds to 
the integration time of the TV cameras used at the ALaMO.

\section{Processing and analysis}
The data from 
ALaMO, FIDAC, VMDB MORP and SSA were separately 
displayed in a crescent cumulative distribution. It was found that
ALaMO, FIDAC, MORP 
and SSA data are best described by a power-law (Equation\ (\ref{powerlaw})). 
Data from VMDB and SPO MORP,
which present a greater magnitude range than the other sources, 
are were better fitted by Equation\ (\ref{qexpum}) 
(details are presented in the following sub-sections 
\ref{sec:meteor-showers} and \ref{sec:lunar-flashes}). 
The parameters were determined with the line search strategy, and the search 
direction was obtained with the conjugate gradient method. 
The adequacy of the fittings, 
as well as the similarity between distributions, 
was established by the Pearson`s' chi-square test. 
All fittings present confidence level equal to or greater than  $95\%$.

\subsection{\label{sec:meteor-showers}Meteor Showers}
\begin{figure}
\begin{center}
 \includegraphics[width=0.9\columnwidth,keepaspectratio,clip]{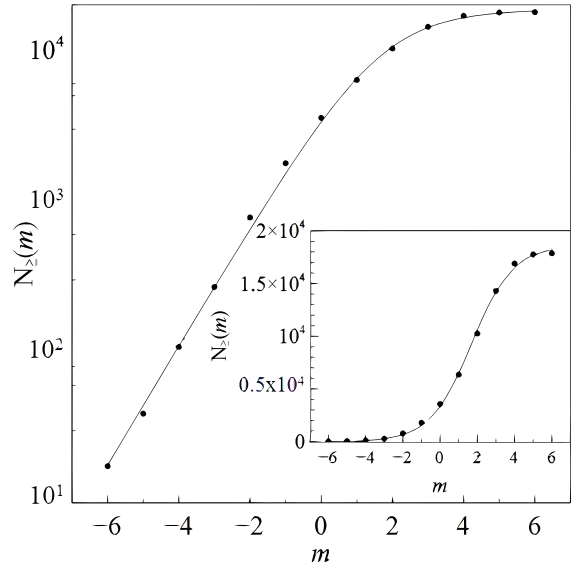}
\end{center}
\caption{\label{fig:per_2000}%
        Cumulative distribution $N_{\geq}(m)$ $\times$ magnitude $m$
        of PER (2000) (circles),
        and fitting with Equation (\protect\ref{qexpum}),
        $N_t=18660$, $\beta'_m= 4.42$, $\gamma=0.879$, $q'= 1.97$.
        Main panel: semi-log scale; inset: linear scale.}
\end{figure}

We have verified that Equation\ (\ref{qexpum}) satisfactorily fits
all VMDB meteor showers, and also the SPO
(see Fig.\ \ref{fig:per_2000}, with PER, that was chosen as a representative 
sample because it is one of the most constant and best observed showers
\citep{beech2004}).
We have obtained mean values of 
$q = 1.57\pm0.05$ ($q'=2.3\pm 0.3$) and $\gamma=1.1\pm0.2$. 
The coincidence of the values of $q$ for different showers suggests that 
the fragmentation process that acts on VMDB meteors are essentially the same 
for the whole sample.
Specifically, the value $q > 1$ implies that  there are other effects 
that influence these distributions besides short-range force.

We have established that $ 97.6\pm  0.5\%$ of VMDB meteors have
magnitudes $m\leq  6$. 
The rest are telescopic meteors. 
The counting of telescopic meteors from PER, ORI, and LEO by \cite{porubcan1973} 
is much smaller than the present prediction. 
This suggests that our result may be taken as an upper bound 
for the number of meteors with $m>6$.  

\begin{figure}
\begin{center}
 \includegraphics[width=0.9\columnwidth,keepaspectratio,clip]{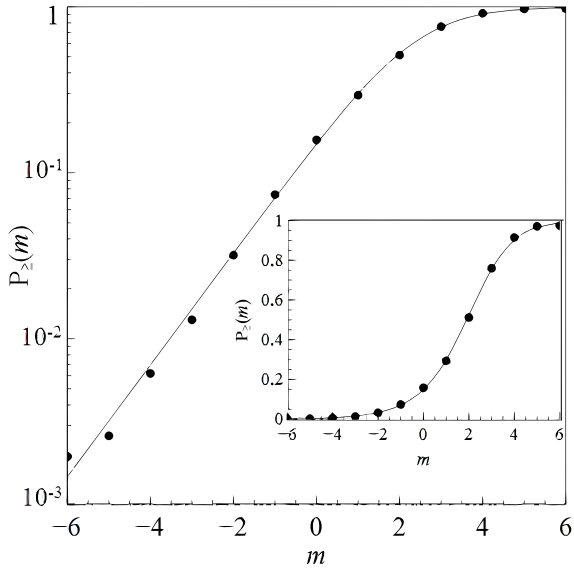}
\end{center}
\caption{\label{fig:capace}%
        Cumulated probability distributions of CAP (2000-2010) (circles) 
        and fitting with Equation (\protect\ref{qexpum})
        with $P_{\ge}(m) = N_{\ge}(m)/N_t$),
        $N_t=1536$, $\beta'_m= 10.20$, $\gamma= 1.13$, $q'= 2.42$ (solid line).
        Main panel: semi-log scale; inset: linear scale.}
\end{figure}

\begin{figure}
\begin{center}
 \includegraphics[width=0.9\columnwidth,keepaspectratio,clip]{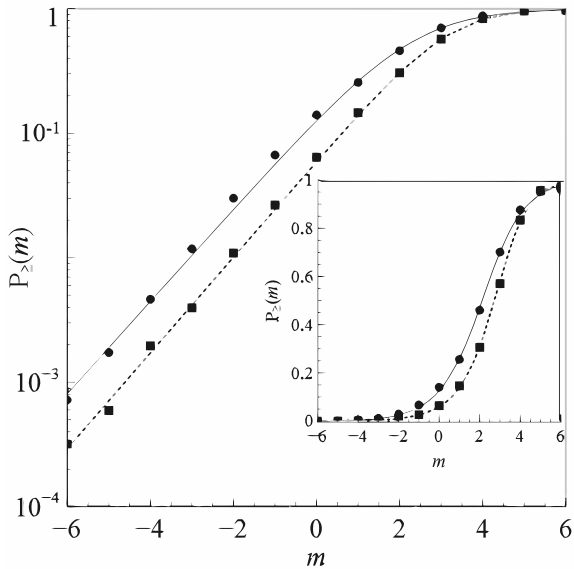}
\end{center}
\caption{\label{fig:cap-spo}%
        Cumulated probability distributions of LEO (1999) (circles) 
        and SPO (2004) (squares), 
        and their corresponding fittings (Eq.~(\protect\ref{qexpum})) with
        $N_t= 60329$, $\beta'_m= 8.72$, $\gamma= 0.99$, $ q'= 2.16$ 
        (LEO, solid line) 
        and $N_t= 21942$, $\beta'_m=48.1$, $\gamma=1.34$, $q'=2.52$ 
        (SPO, dashed line).
}
\end{figure}

We did not detect any temporal variation in the probability distribution of meteors. 
We have compared VMDB data from the year 2000 with the equivalent data
obtained up to the year 2010.  
Specifically for LEO, the counting of meteors  belonging to the 1999 storm 
were compared with those from 2000 to 2010. 
The temporal independence was also verified. 
We have concluded that, despite the possibility of a seasonal increase 
in the total number of meteoroids ($N_t$), the mass distribution in a 
meteor shower remains constant.  
As a consequence of the invariance of the distributions along time,
we found that the meteor distribution may be established using sparse data 
of several years. 
This hypothesis has been tested using all data from 2000 to 2010 of 
the CAP shower,
and they are well adjusted by Equation\ (\ref{qexpum}). 
Probability distribution of the CAP 
is similar to the distributions of other showers
(see Fig.\ \ref{fig:capace}). 
When the showers were separately analysed by their parent bodies,
we find that there are no differences between the meteor distributions 
that originate from 
comets or from asteroids. 
This test was done by comparing the distributions of the ORI and ETA 
(1P/Halley) and GEM (3200 Phaethon) and QUA 
(2003 EH1, \citealt{jenniskens2004}).
The probability distributions of showers associated with comets do not differ 
when they are separated by their dynamic characteristics. 
To this end, we have compared the data showers associated 
with the Halley Family (ETA), the Jupiter Family (LEO), and with 
long period comets (LYR). 
The only difference that we were able to detect 
in 
the VMDB data is between 
SPO meteors and those of all the other showers (see Fig.\ \ref{fig:cap-spo}).

\begin{figure}
\begin{center}
 \includegraphics[width=0.9\columnwidth,keepaspectratio,clip]{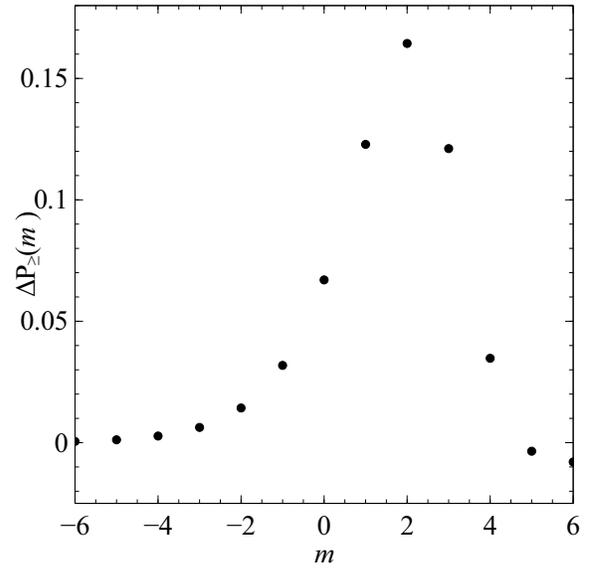}
\end{center}
\caption{\label{fig:difp}%
        Difference between cumulated probabilities of LEO (1999) 
        and SPO (2004), 
${\Delta}P_{\geq}(m)= P_{\geq}^{\text{LEO}}(m)-P_{\geq}^{\text{SPO}}(m)$,
        as a function of the magnitudes $m$,
        putting in evidence the maximum discrepancy at $m=2$.}
\end{figure}

\begin{figure}
\begin{center}
 \includegraphics[width=0.9\columnwidth,keepaspectratio,clip]{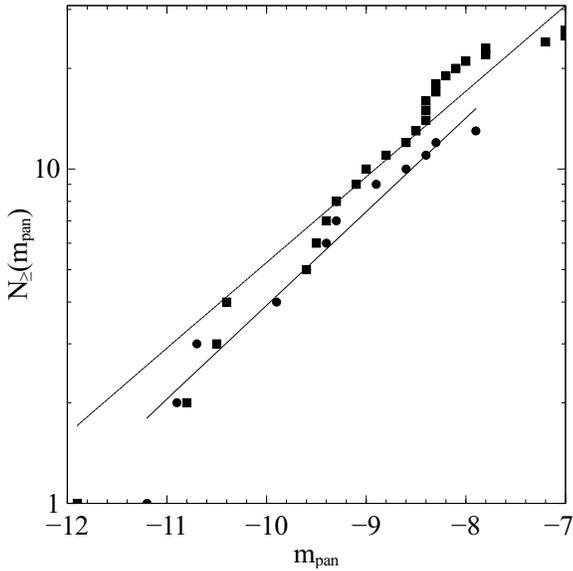}
\end{center}
\caption{\label{fig:morp}%
        Cumulative distribution of the number of meteors MORP 
        $N_{\geq}(m_{\text{{pan}}})$ $\times$ 
        panchromatic magnitude 
        $m_{\text{pan}}$, STA (squares), PER (circles)
        and their corresponding fittings
        (Eq.s~(\ref{powerlaw}) and (\ref{massmagrela})),
        STA (solid line, $n=0.83$) and PER (dashed line, $n= 0.91$). 
        Both fittings using $\gamma= 0.83$ calculated from MORP-SPO data.} 
\end{figure}

The difference between the SPO and shower probability distributions is maximum
for the magnitude 2 (Fig.\ \ref{fig:difp}). In this range, there are 
$\sim 20$\%  more meteors in the showers than observed in SPO.  
The difference is negligible for the magnitudes $-6$ and $+6$.  
This difference may be associated with observational bias
in the process of data collection.  
We have found that most observations of annual showers occur  
$5 \pm 2$ days before or after the peak of the event. 
For instance, in the year 2010, VMDB presents 834 entries
of SPO ($lm \ge 5.5$),
and 434 of those ($\cong 52\%$) correspond to the peak 
period for PER, August, 05--19. 1523 PER entries 
were recorded during this period.
This may be evidence that the sky is not systematically monitored 
except in the annual shower seasons.
The occurrence of observational bias in the VMDB SPO 
data was also suggested when we analysed 
the data from the MORP survey.
The MORP fireballs associated with STA and PER showers were modelled by 
{Equations} (\ref{powerlaw}) and (\ref{massmagrela})  (Fig.\ \ref{fig:morp}).

\begin{figure}
\begin{center}
 \includegraphics[width=0.9\columnwidth,keepaspectratio,clip]{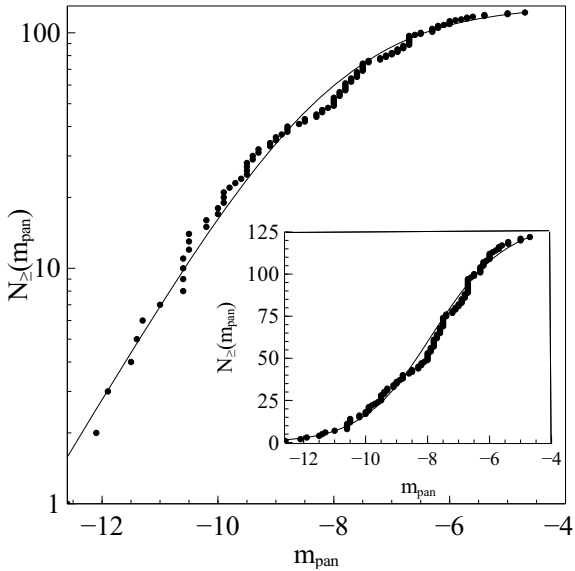}
\end{center}
\caption{\label{fig:morpspo}%
        Cumulative distribution of the number of meteors MORP 
        $N_{\geq}(m_{\text{{pan}}})$ $\times$ $m_{\text{pan}}$,
        MORP-SPO (circles).
        and fitting with Equation (\protect\ref{qexpum}), 
        $N_t=131$, $\beta'_m= 0.0015$, $\gamma= 0.83$, $q'= 1.86$ (solid line).
        Main panel: semi-log scale; inset: linear scale.}
\end{figure}

The comparison between modelled distributions suggested that these showers 
are similar.  
SPO data, however, are modelled by   Equation\ (\ref{qexpum}), 
possibly due to  the magnitude range of this sample 
(Fig. \ref{fig:morpspo}).
The modelled distributions of showers are not correlated to the observed 
distribution of the SPO. 
The number of SPO meteors in the considered magnitude interval is 
systematically greater than that observed in the showers, and this is a 
known result (\citealt{Pawlowski2001}, \citealt{Rendtel2006}).
We obtained  $q=1.51 \pm 0.03$ for the MORP data.
\begin{figure}
\begin{center}
 \includegraphics[width=0.9\columnwidth,keepaspectratio,clip]{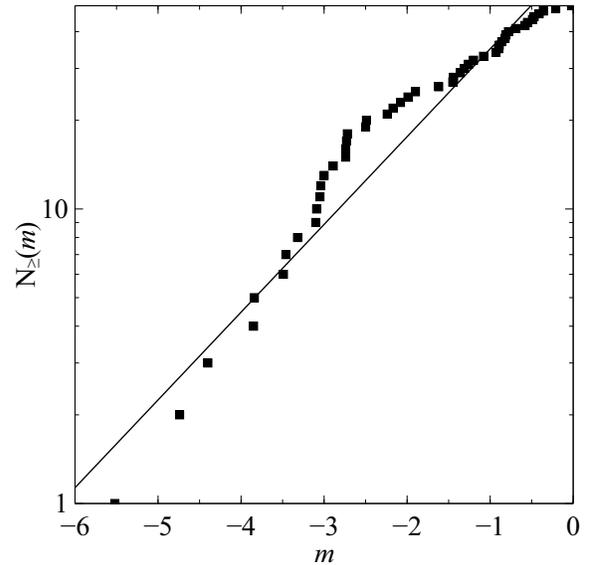}
\end{center}
\caption{\label{fig:rbdm}%
        Cumulative distribution $N_{\geq}(m)$ $\times$ magnitude $m$
        of meteors detected by an all sky camera at Salvador (SSA, squares).
        Solid line represents Equations (\ref{powerlaw}) 
        and (\ref{massmagrela}) ($n=1.2$ e $\gamma=0.59$). 
        The magnitude error is 0.4.
        Fitting is not good for magnitudes around $m=-2$.}
\end{figure}

\begin{figure}
\begin{center}
 \includegraphics[width=0.9\columnwidth,keepaspectratio,clip]{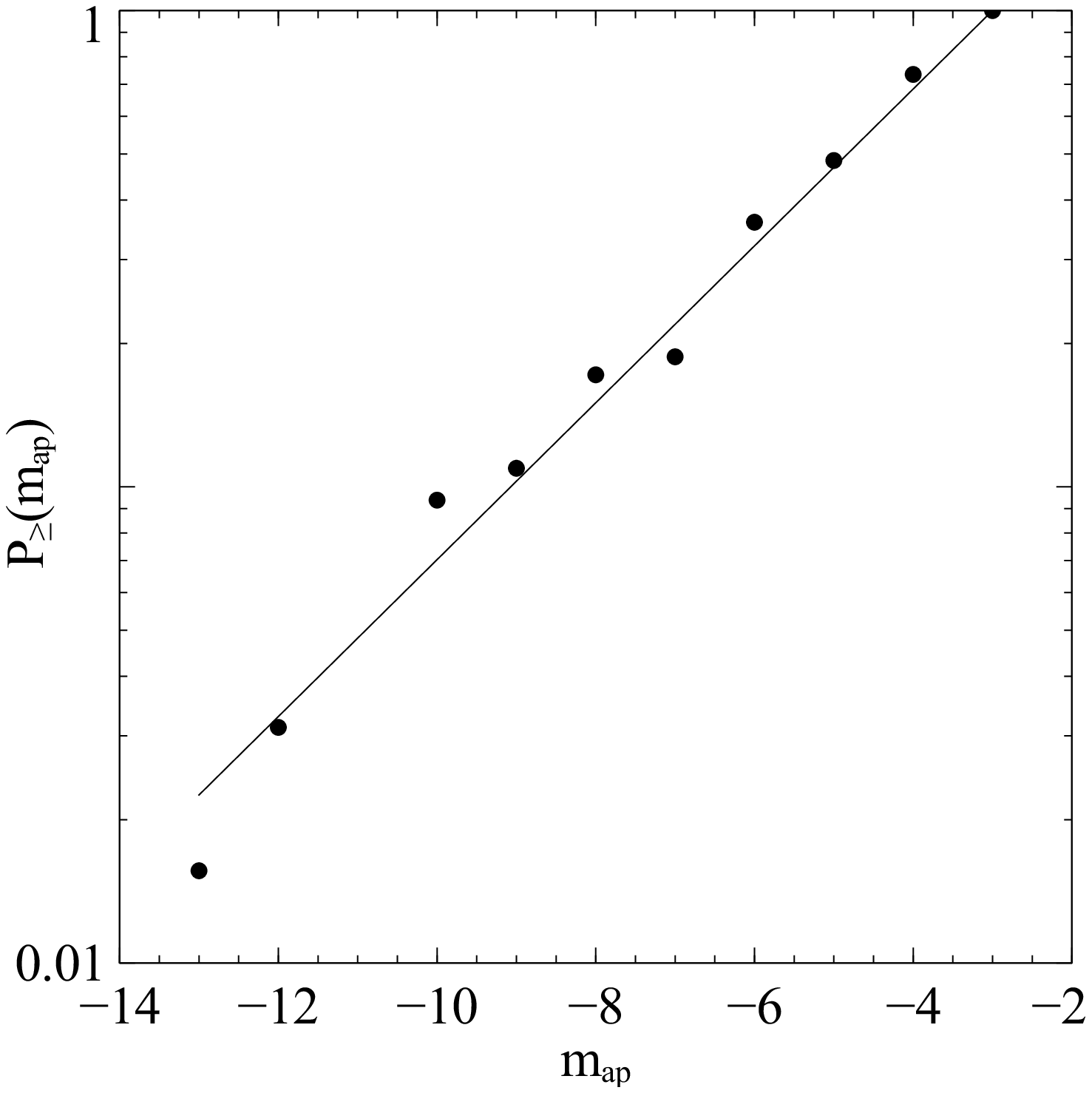}
\end{center}
\caption{\label{fig:fidac}%
        Cumulated probability of FIDAC 
        $P_\geq{(m_{\text{{ap}}})}$ $\times$
        apparent magnitude $m_{\text{ap}}$
        (data from file FIDAC95.001)
        Solid line, Eq.s~(\ref{powerlaw})--(\ref{massmagrela}),
        with $n= 0.33$ and $\gamma= 1.1$.
}
\end{figure}

The SSA data are well fitted with Equations (\ref{powerlaw}) and 
(\ref{massmagrela}). 
The value obtained is $q=1.46 \pm 0.04$ (see Fig. \ref{fig:rbdm}).
The FIDAC's data were also adjusted by Equation (\ref{powerlaw}) 
and  (\ref{massmagrela}), with $q = 1.8\pm0.1$ (Fig.\ \ref{fig:fidac}).
We have found that the FIDAC apparent magnitudes have good 
adherence to the model. 
The same is not true for zenithal magnitude distribution. 
This lack of agreement with the model is associated 
with the transformation of apparent to zenithal magnitudes. 
This conversion takes into account an estimate of the fireball  altitude.
The estimate of this parameter can cause the introduction of an additional 
source of error, as suggested by \cite{bellot1995}.
This problem is particularly important for fireballs brighter than
magnitude $-6$.
We have verified the occurrence of a temporal variation in
the distribution of apparent magnitudes comparing the data from
1993 to 1997.
The compatibility only  occurs when we consider
a range of magnitudes between -6 and -3. 
This suggests the occurrence of  observational bias in the FIDAC sample. 
This bias may be associated with the difference of the terrestrial surface 
area covered by the surveys, total duration of the observations and 
field of view. 
Considering these factors, a compatibility was obtained by \cite{zotkin1978}
comparing visual data of fireballs observed in the former Soviet Union
with meteor photographic networks. For the same reason, this compatibility
also does not occur between the MORP and SSA data. 

\subsection{\label{sec:lunar-flashes}Lunar Flashes}
\begin{figure}
\begin{center}
 \includegraphics[width=0.9\columnwidth,keepaspectratio,clip]{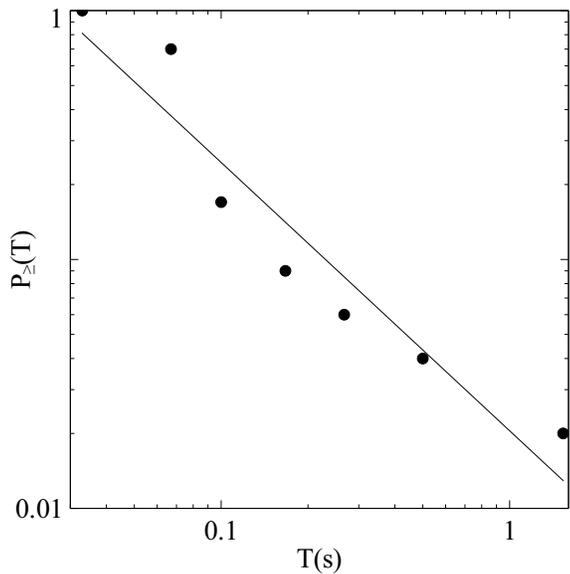}
\end{center}
\caption{\label{fig:alamo}%
        Cumulative probability for the duration $T(s)$ of ALaMO lunar flashes 
        for SPO (circles). 
        The solid line corresponding  the fitting of Eq.~(\ref{powerlaw}) 
        with $n=1.6$.  }
\end{figure}
ALaMO's data are modelled by Equation\ (\ref{powerlaw}) 
for the  impact durations (Fig.\ \ref{fig:alamo}). 
Distributions for the LEO, LYR and GEM showers are similar
(this corresponds to half of the analysed sample),
and this is not observed for the other showers.
On the other hand, it would be expected that SPO distribution differs
from the shower distributions, but we have observed that SPO
distribution present similarities with ORI, GEM, LYR and LEO.
These results disagree with the conclusions obtained 
with the VMDB and MORP data.
This suggests the occurrence of observational bias in the data set,
that can be connected to inadequate observation conditions, 
such as the occurrence of unfavourable Moon phases for the detection of the 
flashes, and/or bad weather. 
Due to bias, it is not possible to infer the existence
of lunar flashes with duration less than 0.033~s.
The mean value for the similar meteor showers is $q=1.32 \pm 0.02$,
and this value is close to the one obtained by photometric data.
Since both phenomena are supposedly directly associated to the
meteoroid masses, we can conclude that the showers distributions observed 
on the Earth and on the  Moon are compatible.
As in the case of meteor showers, we suppose that the mechanisms that govern 
the duration of lunar flashes are both short- and long-range in nature.

\section{Conclusions}

We analysed the distribution of magnitudes of meteors 
and duration of lunar flashes through nonextensive statistics.
We used data from various sources that cover a wide range, 
from telescopic meteors to lunar impactors.
Our main conclusions are as follows:

\begin{enumerate}

\item 

The cumulative distribution of the magnitudes of the 
meteors is well represented by a $q$-exponential. 
This distribution is valid for meteoroids of masses 
varying from telescopic meteors to lunar impact bodies. 
A power-law, that is the asymptotic limit of the $q$-exponential,
is observed for smaller intervals of masses.

\item

We estimate that $2.4 \pm 0.5 \% $ (upper bound limit) 
of meteoroids in meteors showers are telescopic ($m > 6$).

\item

The cumulative distributions of all meteor showers registered 
in VMDB and MORP datasets are similar, 
independent of time, 
independent of the type of parent body (whether asteroid or comet) 
or its dynamic family.

\item
In the distributions of sporadic (non-shower) meteors the VMDB may
present observational bias. 
For instance, the probability of occurrence of meteors of magnitude 2
in showers is about $ 20 \% $ higher than the equivalent in SPO.

\item

The cumulative distribution of duration of lunar flashes 
registered by ALaMO is modelled by a power-law.
The analysis of the distributions suggests that the meteor showers 
observed on the Earth and on the Moon are compatible.

\end{enumerate}

\section*{Acknowledgements}
EPB acknowledges 
the National Institute of Science and Technology for Complex Systems, 
and FAPESB through the program PRONEX (Brazilian agencies).
ASB thanks UFRB/CFP for supporting this work. 
The authors are also grateful to J. Rendtel for suggestions and remarks.
\bibliography{bib}

\begin{thebibliography}{}

\bibitem[\protect\citeauthoryear{{Arlt} \& {Rendtel}}{{Arlt} \&
  {Rendtel}}{2006}]{arlt2006}
{Arlt} R.,  {Rendtel} J.,  2006, MNRAS, 367, 1721

\bibitem[\protect\citeauthoryear{{Baggaley}}{{Baggaley}}{1977}]{baggaley1977}
{Baggaley} W.~J.,  1977, MNRAS, 180, 89

\bibitem[\protect\citeauthoryear{{Beech}, {Illingworth} \& {Brown}}{{Beech}
  et~al.}{2004}]{beech2004}
{Beech} M.,  {Illingworth} A.,    {Brown} P.,  2004, MNRAS, 348, 1395

\bibitem[\protect\citeauthoryear{{Beech}, {Nie} \& {Coulson}}{{Beech}
  et~al.}{2007}]{beech2007}
{Beech} M.,  {Nie} W.,    {Coulson} I.~M.,  2007, JRASC, 101, 139

\bibitem[\protect\citeauthoryear{{Bellot Rubio}}{{Bellot
  Rubio}}{1995}]{bellot1995}
{Bellot Rubio} L.~R.,  1995, A\&A, 301, 602

\bibitem[\protect\citeauthoryear{{Bellot Rubio}, {Ortiz} \& {Sada}}{{Bellot
  Rubio} et~al.}{2000}]{bellot2000}
{Bellot Rubio} L.~R.,  {Ortiz} J.~L.,    {Sada} P.~V.,  2000, EM\&P, 82, 575

\bibitem[\protect\citeauthoryear{{Bernui}, {Tsallis} \& {Villela}}{{Bernui}
  et~al.}{2006}]{2006PhLA..356..426B}
{Bernui} A.,  {Tsallis} C.,    {Villela} T.,  2006, Physics Letters A, 356,
  426, \eprint{astro-ph/0512267},
  \adsurl{http://adsabs.harvard.edu/abs/2006PhLA..356..426B}

\bibitem[\protect\citeauthoryear{{Betzler} \& {Borges}}{{Betzler} \&
  {Borges}}{2012}]{betzler-borges-2012}
{Betzler} A.~S.,  {Borges} E.~P.,  2012, A\&A, 539, A158

\bibitem[\protect\citeauthoryear{{Borges} \& {Tsallis}}{{Borges} \&
  {Tsallis}}{2002}]{epb-tsallis-2002}
{Borges} E.~P.,  {Tsallis} C.,  2002, Phys A, 305, 148,
  \eprint{cond-mat/0109504}

\bibitem[\protect\citeauthoryear{{Bouley}, {Baratoux}, {Vaubaillon}, {Mocquet},
  {Le Feuvre}, {Colas}, {Benkhaldoun}, {Daassou}, {Sabil} \&
  {Lognonn{\'e}}}{{Bouley} et~al.}{2012}]{bouley2012}
{Bouley} S.,  {Baratoux} D.,  {Vaubaillon} J.,  {Mocquet} A.,  {Le Feuvre} M.,
  {Colas} F.,  {Benkhaldoun} Z.,  {Daassou} A.,  {Sabil} M.,    {Lognonn{\'e}}
  P.,  2012, Icar, 218, 115

\bibitem[\protect\citeauthoryear{{Brown} \& {Rendtel}}{{Brown} \&
  {Rendtel}}{1996}]{brown1996}
{Brown} P.,  {Rendtel} J.,  1996, Icar, 124, 414

\bibitem[\protect\citeauthoryear{{Buratti} \& {Johnson}}{{Buratti} \&
  {Johnson}}{2003}]{buratti2003}
{Buratti} B.~J.,  {Johnson} L.~L.,  2003, Icar, 161, 192

\bibitem[\protect\citeauthoryear{{Cardone}, {Leubner} \& {Del
  Popolo}}{{Cardone} et~al.}{2011}]{2011MNRAS.414.2265C}
{Cardone} V.~F.,  {Leubner} M.~P.,    {Del Popolo} A.,  2011, MNRAS, 414, 2265,
  \eprint{1102.3319},
  \adsurl{http://adsabs.harvard.edu/abs/2011MNRAS.414.2265C}

\bibitem[\protect\citeauthoryear{{Ceplecha} \& {Revelle}}{{Ceplecha} \&
  {Revelle}}{2005}]{ceplecha2005}
{Ceplecha} Z.,  {Revelle} D.~O.,  2005, MAPS, 40, 35

\bibitem[\protect\citeauthoryear{{Curado} \& {Tsallis}}{{Curado} \&
  {Tsallis}}{1991}]{curado1991}
{Curado} E.~M.~F.,  {Tsallis} C.,  1991, J. Phys. A, 24, L69

\bibitem[\protect\citeauthoryear{{Davis}}{{Davis}}{2009}]{sanford2009}
{Davis} S.~S.,  2009, Icar, 202, 383

\bibitem[\protect\citeauthoryear{{Dunham}, {Sterner} II, {Gotwols}, {Cudnik},
  {Palmer}, {Sada} \& {Frankenberger}}{{Dunham} et~al.}{2000}]{dunham2000}
{Dunham} D.~W.,  {Sterner} II R.,  {Gotwols} B.,  {Cudnik} B.~M.,  {Palmer}
  D.~M.,  {Sada} P.~V.,    {Frankenberger} R.,  2000, Occultation Newsl., 8, 9

\bibitem[\protect\citeauthoryear{{Halliday}, {Griffin} \&
  {Blackwell}}{{Halliday} et~al.}{1996}]{halliday1996}
{Halliday} I.,  {Griffin} A.~A.,    {Blackwell} A.~T.,  1996, M\&PS, 31, 185

\bibitem[\protect\citeauthoryear{{Hawkes} \& {Jones}}{{Hawkes} \&
  {Jones}}{1986}]{hawkes1986}
{Hawkes} R.~L.,  {Jones} J.,  1986, QJRAS, 27, 569

\bibitem[\protect\citeauthoryear{{Jacchia}, {Verniani} \& {Briggs}}{{Jacchia}
  et~al.}{1965}]{jacchia1965}
{Jacchia} L.~G.,  {Verniani} F.,    {Briggs} R.~E.,  1965, Smithson. Astrophys.
  Obs. Spec. Rep., 175

\bibitem[\protect\citeauthoryear{{Jenniskens}}{{Jenniskens}}{2004}]{jenniskens%
2004}
{Jenniskens} P.,  2004, AJ, 127, 3018

\bibitem[\protect\citeauthoryear{{Jenniskens} \& {Vaubaillon}}{{Jenniskens} \&
  {Vaubaillon}}{2007}]{jenniskens2007}
{Jenniskens} P.,  {Vaubaillon} J.,  2007, AJ, 134, 1037

\bibitem[\protect\citeauthoryear{{Kn{\"o}fel} \& {Rendtel}}{{Kn{\"o}fel} \&
  {Rendtel}}{1988}]{knofel1988}
{Kn{\"o}fel} A.,  {Rendtel} J.,  1988, JIMO, 16, 186

\bibitem[\protect\citeauthoryear{{Koten}}{{Koten}}{1999}]{koten1999}
{Koten} P.,  1999, in {Baggaley} W.~J.,  {Porubcan} V.,  eds, Meteroids 1998
  {Photometry of TV meteors}.
p.~149

\bibitem[\protect\citeauthoryear{Landsberg}{Landsberg}{1990}]{landsberg-1990}
Landsberg P.~T.,  1990, Thermodynamics and Statistical Mechanics.
Dover, New York

\bibitem[\protect\citeauthoryear{{Latora}, {Rapisarda} \& {Tsallis}}{{Latora}
  et~al.}{2001}]{latora-rapisarda-tsallis-2001}
{Latora} V.,  {Rapisarda} A.,    {Tsallis} C.,  2001, Phys. Rev. E, 64, 056134,
  \eprint{cond-mat/0103540}

\bibitem[\protect\citeauthoryear{{Li} \& {Tankin}}{{Li} \&
  {Tankin}}{1987}]{lin1987}
{Li} X.,  {Tankin} R.~S.,  1987, Combust. Sci. and Tech., 56, 65

\bibitem[\protect\citeauthoryear{{Lynden-Bell} \& {Wood}}{{Lynden-Bell} \&
  {Wood}}{1968}]{1968MNRAS.138..495L}
{Lynden-Bell} D.,  {Wood} R.,  1968, MNRAS, 138, 495

\bibitem[\protect\citeauthoryear{{Manson}}{{Manson}}{1995}]{manson1995}
{Manson} J.~W.,  1995, JBAA, 105, 219

\bibitem[\protect\citeauthoryear{{Millman}}{{Millman}}{1980}]{milman1980}
{Millman} P.~M.,  1980, in {Halliday} I.,  {McIntosh} B.~A.,  eds, Solid
  Particles in the Solar System Vol.~90 of IAU Symposium, {One hundred and
  fifteen years of meteor spectroscopy}.
pp 121--127

\bibitem[\protect\citeauthoryear{{Nakamura}, {Fujii}, {Ishiguro}, {Morishige},
  {Yokogawa}, {Jenniskens} \& {Mukai}}{{Nakamura} et~al.}{2000}]{nakamura2000}
{Nakamura} R.,  {Fujii} Y.,  {Ishiguro} M.,  {Morishige} K.,  {Yokogawa} S.,
  {Jenniskens} P.,    {Mukai} T.,  2000, ApJ, 540, 1172

\bibitem[\protect\citeauthoryear{{Ortiz}, {Quesada}, {Aceituno}, {Aceituno} \&
  {Bellot Rubio}}{{Ortiz} et~al.}{2002}]{ortiz2002}
{Ortiz} J.~L.,  {Quesada} J.~A.,  {Aceituno} J.,  {Aceituno} F.~J.,    {Bellot
  Rubio} L.~R.,  2002, ApJ, 576, 567

\bibitem[\protect\citeauthoryear{Padmanabhan}{Padmanabhan}{1990}]{padna2}
Padmanabhan T.,  1990, Phys. Rep., 188, 285

\bibitem[\protect\citeauthoryear{{Pawlowski}, {Hebert}, {Hawkes}, {Matney} \&
  {Stansbery}}{{Pawlowski} et~al.}{2001}]{Pawlowski2001}
{Pawlowski} J.~F.,  {Hebert} T.~J.,  {Hawkes} R.~L.,  {Matney} M.~J.,
  {Stansbery} E.~G.,  2001, M\&PS, 36, 1467

\bibitem[\protect\citeauthoryear{{Porub{\v c}an}}{{Porub{\v
  c}an}}{1973}]{porubcan1973}
{Porub{\v c}an} V.,  1973, BAICz, 24, 1

\bibitem[\protect\citeauthoryear{{Rendtel}}{{Rendtel}}{2006}]{Rendtel2006}
{Rendtel} J.,  2006, JIMO, 34, 71

\bibitem[\protect\citeauthoryear{{Sotolongo-Costa}, {Gamez}, {Luzon}, {Posadas}
  \& {Weigandt Beckmann}}{{Sotolongo-Costa} et~al.}{2007}]{sotolongo2007}
{Sotolongo-Costa} O.,  {Gamez} R.,  {Luzon} F.,  {Posadas} A.,    {Weigandt
  Beckmann} P.,  2007, ArXiv e-prints, \eprint{0710.4963}

\bibitem[\protect\citeauthoryear{{Sotolongo-Costa}, {Grau-Crespo} \&
  {Trallero-Herrero}}{{Sotolongo-Costa} et~al.}{1998}]{sotolongo1998}
{Sotolongo-Costa} O.,  {Grau-Crespo} R.,    {Trallero-Herrero} C.,  1998, Rev.
  Mex. Fis., 44, 441

\bibitem[\protect\citeauthoryear{{Stuart}}{{Stuart}}{1956}]{stuart1956}
{Stuart} L.~H.,  1956, Strolling Astron., 10, 42

\bibitem[\protect\citeauthoryear{{Suggs}, {Cooke}, {Suggs}, {Swift} \&
  {Hollon}}{{Suggs} et~al.}{2008}]{suggs2008}
{Suggs} R.~M.,  {Cooke} W.~J.,  {Suggs} R.~J.,  {Swift} W.~R.,    {Hollon} N.,
  2008, EM\&P, 102, 293

\bibitem[\protect\citeauthoryear{{Thirring}}{{Thirring}}{1970}]{1970ZPhy..235.%
.339T}
{Thirring} W.,  1970, Zeitschrift fur Physik, 235, 339

\bibitem[\protect\citeauthoryear{{Tost}, {Oberst}, {Flohrer} \&
  {Laufer}}{{Tost} et~al.}{2006}]{tost2006}
{Tost} W.,  {Oberst} J.,  {Flohrer} J.,    {Laufer} R.,  2006, in European
  Planetary Science Congress 2006 {Lunar Impact Flashes: History of
  observations and recommendations for future campaigns}.
p.~546

\bibitem[\protect\citeauthoryear{{T{\'o}th} \& {Kla{\v c}ka}}{{T{\'o}th} \&
  {Kla{\v c}ka}}{2004}]{toth2005}
{T{\'o}th} J.,  {Kla{\v c}ka} J.,  2004, EM\&P, 95, 181

\bibitem[\protect\citeauthoryear{{Tsallis}}{{Tsallis}}{1988}]{tsallis1988}
{Tsallis} C.,  1988, J. Stat. Phys., 52, 479

\bibitem[\protect\citeauthoryear{Tsallis}{Tsallis}{1994}]{tsallis-quimicanova1%
994}
Tsallis C.,  1994, {Q}uim. {N}ova, 17, 468

\bibitem[\protect\citeauthoryear{{Tsallis}}{{Tsallis}}{2009}]{tsallis2009}
{Tsallis} C.,  2009, Introduction to Nonextensive Statistical
  Mechanics:Approaching a Complex World.
Springer, New York

\bibitem[\protect\citeauthoryear{{Tsallis}, {Mendes} \& {Plastino}}{{Tsallis}
  et~al.}{1998}]{tsallis1998}
{Tsallis} C.,  {Mendes} R.,    {Plastino} A.~R.,  1998, Phys A, 261, 534

\bibitem[\protect\citeauthoryear{{Tsallis}, {Rapisarda}, {Pluchino} \&
  {Borges}}{{Tsallis} et~al.}{2007}]{tsallis-rapisarda-pluchino-borges-2007}
{Tsallis} C.,  {Rapisarda} A.,  {Pluchino} A.,    {Borges} E.~P.,  2007, Phys
  A, 381, 143, \eprint{cond-mat/0609399}

\bibitem[\protect\citeauthoryear{{Whipple}}{{Whipple}}{1983}]{whipple1983}
{Whipple} F.~L.,  1983, IAUC, 3881, 1

\bibitem[\protect\citeauthoryear{{Zolensky}, {Bland}, {Brown} \&
  {Halliday}}{{Zolensky} et~al.}{2006}]{zolensky2006}
{Zolensky} M.,  {Bland} P.,  {Brown} P.,    {Halliday} I.,  2006, {Flux of
  Extraterrestrial Materials}.
pp 869--888

\bibitem[\protect\citeauthoryear{{Zotkin} \& {Khotinok}}{{Zotkin} \&
  {Khotinok}}{1978}]{zotkin1978}
{Zotkin} I.~T.,  {Khotinok} R.~L.,  1978, Meteoritika, 37, 37

\end{thebibliography}

\bsp
\label{lastpage}
\end{document}